# On the inelastic collapse of a ball bouncing on a randomly vibrating platform

Satya N Majumdar

*Laboratoire de Physique Théorique et Modèles Statistique, Université Paris-Sud.*
*Bât. 100 91405, Orsay Cedex, France*
majumdar@lptms.u-psud.fr

Michael J Kearney

*School of Electronics and Physical Sciences, University of Surrey,*
*Guildford, Surrey, GU2 7XH, United Kingdom*
m.j.kearney@surrey.ac.uk

A theoretical study is undertaken of the dynamics of a ball which is bouncing inelastically on a randomly vibrating platform, motivated as a model problem of inelastic collapse. Of principal interest are the distributions of the number of flights $n_f$ till the collapse and the total time $\tau_c$ elapsed before the collapse. In the strictly elastic case, both distributions have power law tails characterised by exponents which are universal, i.e. independent of the detail of the platform noise distribution. In the inelastic case, both distributions have exponential tails: $P(n_f) \sim \exp[-\theta_1 n_f]$ and $P(\tau_c) \sim \exp[-\theta_2 \tau_c]$. The decay exponents $\theta_1$ and $\theta_2$ depend continuously on the coefficient of restitution and are nonuniversal; however, as one approaches the elastic limit, they vanish in a manner which turns out to be universal. An explicit expression for $\theta_1$ is provided for a particular case of the platform noise distribution.

# I. INTRODUCTION

The study of granular materials and granular flows, see e.g. [1-3], has highlighted the importance of the concept of inelastic collapse [4,5]. The canonical example of inelastic collapse concerns a ball bouncing on a static platform in a constant gravitational field $g$. If $u_0$ is the initial velocity with which the ball is thrown up from the platform, the velocity at subsequent bounces decreases geometrically until the ball stops completely after a time $\tau_c = 2u_0/[g(1-r)]$, where $r$ is the coefficient of restitution. Note that although the total number of flights until the collapse $n_f$ is *infinite*, the total flight time until the collapse is actually *finite*. Analogous collapse transitions are also predicted theoretically for a ball bouncing on a periodically vibrating plate [6-8] and for a randomly accelerated particle near an absorbing boundary [9-13]. Elegant experimental studies of inelastic collapse in granular layers subject to vertical vibrations may be found in [14,15].

The question we address in this paper is as follows: How do these two physical observables, namely (i) the number of flights $n_f$ till the collapse and (ii) the total time $\tau_c$ elapsed before the collapse, behave for a ball bouncing on a platform which itself vibrates in a *noisy* manner? Let us first see what happens in a single collision between the ball and the platform. Let $\upsilon$ be the velocity of the platform and $-u_i$ be the incident velocity of the ball. Then, just after the collision, the platform velocity remains unchanged at $\upsilon$ but the ball bounces with velocity

$$u_b = ru_i + (1+r)\upsilon \qquad (1)$$

where the coefficient of restitution $0 \leq r \leq 1$ and $r = 1$ corresponds to the elastic limit. This result can easily be derived by considering an inelastic collision between two particles of mass $m_1$ and $m_2$ with incident velocities $u_1$ and $u_2$ respectively. Let $u_1'$ and $u_2'$ be the post-collision velocities. Conservation of momentum implies that $m_1 u_1 + m_2 u_2 = m_1 u_1' + m_2 u_2'$ and inelasticity implies $u_2' - u_1' = -r(u_2 - u_1)$. Solving these two equations when $m_2 \gg m_1$ one gets $u_2' = u_2$ and $u_1' = -r u_1 + (1+r) u_2$. In our example, the platform corresponds to the massive particle with velocity $u_2 = v$ and the incident velocity of the ball is $u_1 = -u_i$. This gives the result in Eq. (1). Thus, the bounce velocity $u_n$ of the ball after the $n$-th collision with the platform satisfies a simple recurrence relation

$$u_n = r u_{n-1} + \eta_n \qquad (2)$$

where $\eta_n = (1+r) v_n$ with $v_n$ being the velocity of the platform at the time of the $n$-th collision. The recursion in Eq. (2) starts with initial value $u_0 > 0$.

We take the simplest model of a noisy platform, which is broadly applicable if $u_n \gg |v_n|$, by assuming that the velocities of the platform at different collision times are completely uncorrelated and each of them is drawn independently from a given distribution. This is equivalent to saying that the noise variables $\eta_n$ in Eq. (2) are independent and identically distributed random variables, each drawn from a given distribution $\rho(\eta)$. We further assume that $\rho(\eta)$ is continuous and symmetric around $\eta = 0$, i.e. $\langle \eta_n \rangle = 0$, and we scale the noise so that it has unit variance, i.e. $\langle \eta_n^2 \rangle = 1$. Thus the typical platform velocity during a collision $|v_n| \sim O(1)$. In practice, once $u_n \sim O(1)$ there will be an increased likelihood that the platform velocity at impact will be positive [14] and the above assumptions break down. However, in such circumstances we will assume that the ball basically 'sticks' to the platform through a process known as *chattering* [7]. Hence, in what follows, noting that once the bounce velocity becomes negative *for the first time* the ball effectively hasn't bounced, we will take this *first-passage* event as the definition of collapse.

When the platform has noisy vibrations both the number of flights $n_f$ and the total time $\tau_c$ till the collapse become random variables. Our main results, summarized below, concern the distributions $P(n_f)$ and $P(\tau_c)$ of these two random variables.

(i)  In the strictly elastic case $r = 1$, we show that both of these distributions have power law tails: $P(n_f) \sim n_f^{-\alpha_1}$ for large $n_f$ and $P(\tau_c) \sim \tau_c^{-\alpha_2}$ for large $\tau_c$, where the exponents $\alpha_1 = 3/2$ and $\alpha_2 = 4/3$ are universal, i.e., independent of the noise distribution $\rho(\eta)$, as long as it is continuous and symmetric.

(ii)  In the inelastic case $r < 1$, on the other hand, both of these distributions have exponential tails: $P(n_f) \sim \exp[-\theta_1(r) n_f]$ for large $n_f$ and $P(\tau_c) \sim \exp[-\theta_2(r) \tau_c]$ for large $\tau_c$. The decay exponents $\theta_1(r)$ and $\theta_2(r)$ depend continuously on $r$ and, moreover, they are nonuniversal as they depend explicitly on the noise distribution $\rho(\eta)$. The exact expressions for $\theta_1(r)$ and $\theta_2(r)$ for an arbitrary noise distribution $\rho(\eta)$ are, in general, hard to obtain. As an illustration, we provide in an Appendix explicit results for $\theta_1(r)$ for the special noise distribution, $\rho(\eta) = \frac{1}{2}\exp[-|\eta|]$.

(iii)    As one approaches the elastic limit, $r \to 1$, the decay exponents $\theta_1(r)$ and $\theta_2(r)$ vanish irrespective of the noise distribution. Most interestingly, the manner in which these decay exponents approach zero as $r \to 1$ turns out to be universal. For example, we find that as $r \to 1$, $\theta_1(r) \approx a(1-r)$ and $\theta_2(r) \approx bg(1-r)^{3/2}$ where $g$ is the gravitational constant and $a, b$ are universal constants independent of the noise distribution $\rho(\eta)$. We show that while $a = 1$ is trivial, $b = 0.405024...$ is nontrivial and is given by the smallest positive root of the equation, $D_{2b^2}[-2\sqrt{2}b] = 0$ where $D_p(z)$ is the parabolic cylinder function of index $p$ and argument $z$ [16].

## II. STRICTLY ELASTIC CASE

We first consider the strictly elastic case $r = 1$, against which we wish to make comparison once we have considered the inelastic case $r < 1$. The recursion relation in Eq. (2) with $r = 1$ reduces to a standard random walk

$$u_n = u_{n-1} + \eta_n \tag{3}$$

starting with the initial value $u_0 > 0$. The noise $\eta_n$ is drawn independently for each $n$ from the distribution $\rho(\eta)$ such that $\langle \eta_n \rangle = 0$ and $\langle \eta_n^2 \rangle = 1$. We would like to compute the probability distributions of (i) the number of flights or collisions $n_f$ before the collapse, i.e., before the velocity $u_n$ evolving via the random sequence in Eq. (3) becomes negative for the first time, and (ii) the total flight time $\tau_c = \frac{2}{g}\sum_{n=0}^{n_f} u_n$ elapsed before the collapse.

These two distributions are, in general, nonuniversal in the sense that they depend explicitly on the noise distribution $\rho(\eta)$ and are hard to compute for generic $\rho(\eta)$. For example, let us consider how one would calculate the distribution of the number of collisions before the collapse, $P(n_f, u_0)$. Define $Q(n, u_0)$ to be the probability that the sequence in Eq. (3), starting at $u_0$, does not change sign unto step $n$. Then $P(n_f, u_0) = Q(n_f - 1, u_0) - Q(n_f, u_0)$. The probability $Q(n, u_0)$ satisfies the following integral equation,

$$Q(n, u_0) = \int_0^\infty Q(n-1, u') \rho(u' - u_0) du' \qquad (4)$$

starting from the initial condition, $Q(0, u_0) = 1$ for all $u_0 > 0$. Eq. (4) follows from considering what happens at the first step: suppose the ball jumps from $u_0$ to $u' > 0$, with probability $\rho(u' - u_0)$. For the subsequent $(n-1)$ steps, the probability of no sign change, starting initially at $u'$, is just $Q(n-1, u')$. The range of the integral $[0, \infty]$ for $u'$ in Eq. (4) ensures that the process didn't change sign in the first step. Note that Eq. (4) uses the Markov property of the sequence in Eq. (3). Even though the right hand side of Eq. (4) has a convolution form, the exact solution for arbitrary $n$ is difficult due to the fact that the lower limit of the integral is zero. Naturally, the distribution of the collapse time $\tau_c$ will be even harder to compute analytically.

There is, however, an important theorem due to Sparre Andersen [17] which, in relation to the sequence Eq. (3), states the following: Given initial condition $u_0 = 0$, the probability $P(n)$ that the process first becomes negative on the $n$-th step is actually *independent* of the step distribution $\rho(\eta)$, as long as it is continuous and symmetric. In fact, the theorem proves that $P(n) = \Gamma(n - \frac{1}{2})/[2\sqrt{\pi} n!]$, which implies that $P(n) \sim n^{-3/2}$ for large $n$ [18]. In our problem $u_0 > 0$, so this theorem does not strictly apply; nevertheless, it makes itself felt when discussing the tail of the distribution of the number of collisions before the collapse, $P(n_f)$. We show later that whilst the distributions $P(n_f)$ and $P(\tau_c)$ are generically nonuniversal, their tails are algebraic with exponents that are universal. The behaviour at the tails can be derived by analysing the corresponding quantities for a continuous time Brownian motion, which is much easier to analyse. Therefore, we first review these quantities for the continuous time Brownian motion. In subsection II.B, we use these results to predict the tails of the distributions in the discrete case.

### A. Continuous time Brownian motion

Consider a Brownian motion $y(t)$ evolving in continuous time $t$ via the Langevin equation

$$\frac{dy}{dt} = \xi(t) \tag{5}$$

where $\xi(t)$ is a zero mean white noise with correlator $\langle \xi(t)\xi(t')\rangle = \delta(t-t')$. The motion starts at $y(t=0) = y_0$. It is relatively straightforward to compute the probability distributions of (i) the first-passage time $t_f$ at which the process crosses zero for the first time, and (ii) the area swept out by the process *till its first-passage time*, $A = \int_0^{t_f} y(t')\,dt'$. Following [19], we first consider (for reasons that become clear later) the probability distribution $P(T, y_0)$ of the general observable

$$T = \int_0^{t_f} V[y(t')]\,dt' \tag{6}$$

where $V[y(t)]$ is an arbitrary functional of the process $y(t)$ and $t_f$ is the first-passage time of the process. It is useful to study the Laplace transform of this distribution,

$$\tilde{P}(s, y_0) = \left\langle \exp\left[-s\int_0^{t_f} V[y(t')]\,dt'\right]\right\rangle = \int_0^{\infty} P(T, y_0) e^{-sT}\,dT \tag{7}$$

where $\langle\ \rangle$ denotes an average over all realizations of the process till its first-passage time. A typical path of the process over the time interval $[0, t_f]$ may be split into an initial step $\Delta y$ during the interval $[0, \Delta t]$ followed by the remaining steps during the interval $[\Delta t, t_f]$. Then Eq. (7) can be written as

$$\tilde{P}(s, y_0) = \left\langle \exp\left[-s\int_0^{t_f} V[y(t')]\,dt'\right]\right\rangle = \left\langle e^{-sV[y_0]\Delta t}\,\tilde{P}(s, y_0 + \Delta y)\right\rangle_{\Delta y} \tag{8}$$

where $\Delta y = \xi(0)\Delta t$. The average in the second part of Eq. (8) is over all possible realizations of $\Delta y$. Expanding in powers of $\Delta t$ and using the fact that the noise $\xi(t)$ is delta correlated, i.e., $\langle \xi(0)^2 \rangle \approx 1/\Delta t$ as $\Delta t \to 0$, one gets a backward Fokker-Planck equation for $\tilde{P}(s, y_0)$ in the space of the initial position $y_0$,

$$\left[ \frac{1}{2} \frac{\partial^2}{\partial y_0^2} - sV[y_0] \right] \tilde{P}(s, y_0) = 0. \tag{9}$$

This differential equation is valid in the range $y_0 \in [0, \infty]$ and satisfies the elementary boundary conditions $\tilde{P}(s, y_0 \to 0) = 1$ and $\tilde{P}(s, y_0 \to \infty) = 0$.

We now consider certain choices for $V[y]$ to make a connection with the problems of interest. For the first-passage time distribution we need to choose $V[y] = 1$, so that $T = t_f$ from Eq. (6). The solution of Eq. (9) with $V[y_0] = 1$ can be easily obtained,

$$\tilde{P}(s, y_0) = \exp(-\sqrt{2s}\, y_0). \tag{10}$$

Inverting the Laplace transform, we get the required first-passage time distribution, a result which is well-known, see e.g. [20,21],

$$P(t_f, y_0) = \frac{1}{\sqrt{2\pi}} \frac{y_0}{t_f^{3/2}} \exp\left( -\frac{y_0^2}{2t_f} \right) \tag{11}$$

valid for all $y_0 > 0$ and $t_f > 0$. Of primary importance is the fact that in the limit $t_f \gg y_0^2$, this distribution has an algebraic tail,

$$P(t_f, y_0) \approx \frac{1}{\sqrt{2\pi}} \frac{y_0}{t_f^{3/2}}. \tag{12}$$

Regarding the distribution of the area $A = \int_0^{t_f} y(t') dt'$ swept under a Brownian curve till its first-passage time, we choose $V[y] = y$ in Eq. (6) so that $T = A$. Then Eq. (9) with $V[y_0] = y_0$ has the solution

$$\tilde{P}(s, y_0) = \frac{2^{7/6} s^{1/6}}{\Gamma(\tfrac{1}{3}) 3^{1/3}} \sqrt{y_0} K_{1/3}\left(\sqrt{\frac{8 s y_0^3}{9}}\right). \tag{13}$$

where $K_{1/3}(z)$ is a modified Bessel function [16]. The Laplace transform in Eq. (13) can be inverted to give the distribution $P(A, y_0)$ for all $A > 0$ and all $y_0 > 0$ [19,21]

$$P(A, y_0) = \frac{2^{1/3}}{3^{2/3} \Gamma(\tfrac{1}{3})} \frac{y_0}{A^{4/3}} \exp\left(-\frac{2 y_0^3}{9 A}\right). \tag{14}$$

Again, for $A \gg y_0^3$ this distribution has an algebraic tail

$$P(A, y_0) \approx \frac{2^{1/3}}{3^{2/3} \Gamma(\tfrac{1}{3})} \frac{y_0}{A^{4/3}}. \tag{15}$$

Although the above results are relatively standard, the backward Fokker-Planck technique used will prove useful in the discussion of the inelastic case in section III. The same method was also employed in the context of an undamped particle moving in a random Sinai potential [22], and other examples may be found in [23].

## B. The relationship between the discrete sequence and the continuous Brownian motion

As mentioned above, the tails of various distributions in the discrete case can be obtained by analysing their continuous counterparts. To make the connection between the discrete and the continuous case, let us evolve the discrete sequence in Eq. (3) up to step $n_0 \gg 1$, ensuring that it didn't change sign in between. At step $n_0$, the typical value of $u_{n_o} \sim \sqrt{n_0} \gg O(1)$, a fact that simply follows from the central limit theorem. Let us now consider further evolution of the sequence for $n > n_0$, for which while $u_n \gg O(1)$, the increment $\delta u_n = \eta_n \sim O(1)$. We define a new scaled variable, $y_n = u_n / \sqrt{n_0}$ which then evolves for $n > n_0$ as

$$y_n = y_{n-1} + \frac{1}{\sqrt{n_0}} \eta_n \tag{16}$$

starting at $y_{n_0} = u_{n_0} / \sqrt{n_0} \sim O(1)$ at $n = n_0$. We next define $\Delta t = 1/n_0$. In the limit of large $n_0$, $\Delta t$ becomes small and $t = (n - n_0)\Delta t$ can be considered as a continuous 'time' variable. Writing $y_n \equiv y(t = (n - n_0)\Delta t)$, dividing the Eq. (16) by $\Delta t = 1/n_0$ and taking the limit of large $n_0$, we get back Eq. (5) for the continuous time Brownian motion following the identification $\xi(t) = \eta_n / \sqrt{\Delta t} = \sqrt{n_0} \eta_n$. Thus, for $n > n_0$ (where $n_0$ is large), we expect that the results for the discrete sequence and the continuous process will coincide.

For example, the probability distribution $P(n_f, n_0)$ of the number of flights $n_f$ before the collapse in the discrete sequence, for $n_f > n_0$, can be read off the corresponding continuous result in Eq. (11) after making the substitutions $y_{n_0} = u_{n_0} / \sqrt{n_0}$ and $t_f = (n_f - n_0)\Delta t = (n_f - n_0)/n_0$, and we get

$$P(n_f, u_{n_0}) \approx \frac{1}{\sqrt{2\pi}} \frac{u_{n_0}}{(n_f - n_0)^{3/2}} \exp\left(-\frac{u_{n_0}^2}{2(n_f - n_0)}\right). \tag{17}$$

This distribution thus has a power law tail for $n_f \gg n_0$

$$P(n_f, u_{n_0}) \approx \frac{1}{\sqrt{2\pi}} \frac{u_{n_0}}{n_f^{3/2}}. \tag{18}$$

This is the expression for the probability distribution of $n_f$ given that the sequence didn't change sign up to $n_0 \gg 1$ and that the value of the sequence at step $n_0$ is $u_{n_0}$. To calculate the unconditional distribution, one then has to average the expression in Eq. (18) over $u_{n_0}$,

$$P(n_f) = \langle P(n_f, u_{n_0}) \rangle_{u_{n_0}} \approx \frac{1}{\sqrt{2\pi}} \frac{\langle u_{n_0} \rangle}{n_f^{3/2}}. \tag{19}$$

The exponent 3/2 comes from the continuous case and hence is universal, i.e., independent of the noise distribution $\rho(\eta)$ of the discrete sequence. Naturally, this observation is intimately related to the Sparre Andersen theorem discussed earlier. Note, however, that the amplitude of the power law decay contains $\langle u_{n_0} \rangle$ which depends explicitly on the noise distribution $\rho(\eta)$.

In a similar way one can also derive, for the discrete sequence, the behaviour at the tail of the distribution $P(\tau_c)$ of the total time elapsed till the collapse, $\tau_c = \frac{2}{g} \sum_{n=0}^{n_f} u_n$. We split this sum into two parts, $\tau_c = \tau_{n_0} + S$, where $\tau_{n_0}$ is the time elapsed up to step $n_0 \gg 1$ (given that the sequence didn't change sign in between) and $S = \frac{2}{g} \sum_{n_0}^{n_f} u_n$ is the time elapsed from step $n_0$ up to the collapse at step $n_f$. For large $\tau_c$, $S \gg \tau_{n_0}$ and hence $\tau_c \approx S$. So, to determine the distribution of $\tau_c$ for large $\tau_c$, we need to calculate the distribution $P(S, u_{n_0})$ of $S$, given $u_{n_0}$. Following the argument outlined before, the distribution $P(S, u_{n_0})$ can be directly read off from Eq. (14) after the substitutions $t_f = (n_f - n_0)\Delta t = (n_f - n_0)/n_0$ and $y_n \equiv y(t) = u_n/\sqrt{n_0}$, after first noting that

$$S = \frac{2}{g} \sum_{n=n_0}^{n_f} u_n = \frac{2}{g} \sqrt{n_0} \sum_{n=n_0}^{n_f} y_n \approx \frac{2}{g} n_0^{3/2} \int_0^{t_f} y(t')\,dt' = \frac{2}{g} n_0^{3/2} A. \tag{20}$$

Using Eq. (14), we then get

$$P(S, u_{n_0}) \approx c_1 \frac{u_{n_0}}{S^{4/3}} \exp\left(-\frac{4u_{n_0}^3}{9gS}\right) \tag{21}$$

where $c_1 = \left(\frac{2}{3}\right)^{2/3} / [\Gamma(\frac{1}{3}) g^{1/3}]$. For large $S$ this distribution has an algebraic tail

$$P(S, u_{n_0}) \approx c_1 \frac{u_{n_0}}{S^{4/3}}. \tag{22}$$

Averaging over $u_{n_0}$ and using $\tau_c \approx S$ for large $\tau_c$, we get

$$P(\tau_c) \approx c_1 \frac{\langle u_{n_0} \rangle}{\tau_c^{4/3}}. \tag{23}$$

Once again, the exponent 4/3 is universal since it comes from the continuous Brownian motion and is independent of the noise distribution $\rho(\eta)$.

Noting that the same nonuniversal quantity $\langle u_{n_0} \rangle$ appears in the amplitude of the tails of both the distributions $P(n_f)$ in Eq. (19) and $P(\tau_c)$ in Eq. (23), we find that the amplitude ratio,

$$\zeta = \frac{\lim_{\tau_c \to \infty} \tau_c^{4/3} P(\tau_c)}{\lim_{n_f \to \infty} n_f^{3/2} P(n_f)} = \left(\frac{2}{3}\right)^{2/3} \frac{\sqrt{2\pi}}{[\Gamma(\frac{1}{3}) g^{1/3}]} \tag{24}$$

is a universal number independent of the noise distribution $\rho(\eta)$, providing that this distribution is continuous and symmetric.

## III. INELASTIC CASE

We now turn to the sequence in Eq. (2) with $0 < r < 1$. As before, the sequence starts from the initial value $u_0 > 0$ and we want to compute two quantities: (i) the distribution $P(n_f)$ of the number of flights before the collapse, i.e., before the velocity becomes negative for the first time, and (ii) the distribution $P(\tau_c)$ of the total time $\tau_c = \frac{2}{g} \sum_{n=0}^{n_f} u_n$ elapsed before the collapse.

Unlike the elastic case $r = 1$, these two distributions are generically nonuniversal for $r < 1$ and depend quite explicitly on the noise distribution $\rho(\eta)$ [18]. The probability $Q(n, u_0)$ that the sequence does not become negative up to step $n$, starting with $u_0$ at $n = 0$, satisfies the integral equation

$$Q(n, u_0) = \int_0^\infty Q(n-1, u') \rho(u' - r u_0) du' \qquad (25)$$

with the initial condition, $Q(0, u_0) = 1$ for all $u_0 > 0$. This equation appeared recently in the context of the persistence of a continuous stochastic process with discrete time sampling [24]. There, it was shown that $Q(n, u_0) \sim \exp[-\theta_1(r) n]$ for large $n$, where the decay exponent $\theta_1(r)$ depends continuously on $r$ and also on the form of the noise distribution $\rho(\eta)$. Thus, $P(n_f) \sim \exp[-\theta_1(r)n_f]$ for large $n_f$. For the special case of Gaussian noise $\rho(\eta) = \frac{1}{\sqrt{2\pi}}\exp(-\eta^2/2)$, various methods have been developed to determine the decay constant $\theta_1(r)$ accurately, though an exact explicit solution is still missing [24]. In an Appendix, we show how to explicitly determine $\theta_1(r)$ for the exponentially distributed noise $\rho(\eta) = \frac{1}{2}\exp(-|\eta|)$, although the analysis is far from easy. The calculation of the collapse time distribution $P(\tau_c)$ is even harder. However, one expects that $P(\tau_c)$ also has a similar exponential tail, $P(\tau_c) \sim \exp[-\theta_2(r)\tau_c]$ for large $\tau_c$ where the decay exponent $\theta_2(r)$ is also nonuniversal.

In the elastic case $r = 1$, both $P(n_f)$ and $P(\tau_c)$ have power law tails. This means that both the decay exponents $\theta_1(r)$ and $\theta_2(r)$, characterizing the exponential tails for $r < 1$, must vanish as one approaches the elastic limit $r \to 1$. We show later in this section that these two exponents vanish in a universal fashion in the limit $r \to 1$: $\theta_1(r) \approx a(1-r)$ and $\theta_2(r) \approx bg(1-r)^{3/2}$, where $a = 1$ and $b = 0.405024...$ are universal constants independent of the noise distribution $\rho(\eta)$. The reason for this universality can be traced to the fact that in the limit $r \to 1$, the discrete sequence in Eq. (2) can be approximated by a continuous time Ornstein-Uhlenbeck process.

### A. Continuous time Ornstein-Uhlenbeck process

Consider a stochastic process $y(t)$ evolving via the continuous time Ornstein-Uhlenbeck (OU) equation

$$\frac{dy}{dt} = -\lambda y + \xi(t) \tag{26}$$

where $\xi(t)$ is a zero mean white noise with correlator $\langle \xi(t)\xi(t')\rangle = \delta(t-t')$ and the motion starts at $y(t=0) = y_0$. This represents the Langevin equation of a particle moving in a parabolic potential. We compute below the probability distributions of the two quantities of interest: (i) the first-passage time $t_f$ at which the process first crosses zero and (ii) the area swept out by the process till its first-passage time, $A = \int_0^{t_f} y(t')dt'$.

As was the case with the ordinary Brownian motion studied in subsection II.A, we first consider the general case where we compute the probability distribution $P(T, y_0)$ of the observable $T = \int_0^{t_f} V[y(t')]dt'$. Using the same backward Fokker-Planck technique, it is straightforward to show that the Laplace transform $\tilde{P}(s, y_0)$ satisfies the differential equation in $y_0$,

$$\left[ \frac{1}{2} \frac{\partial^2}{\partial y_0^2} - \lambda y_0 \frac{\partial}{\partial y_0} - sV[y_0] \right] \tilde{P}(s, y_0) = 0 \tag{27}$$

with the boundary conditions $\tilde{P}(s, y_0 \to 0) = 1$ and $\tilde{P}(s, y_0 \to \infty) = 0$. To calculate the first-passage time distribution, we choose $V[y] = 1$, so that $T = t_f$. With $V[y_0] = 1$, the exact solution of Eq. (27) is given by,

$$\tilde{P}(s, y_0) = e^{\lambda y_0^2 / 2} \frac{D_{-s/\lambda}(y_0 \sqrt{2\lambda})}{D_{-s/\lambda}(0)} \tag{28}$$

where $D_p(z)$ is the parabolic cylinder function of index $p$ and argument $z$ [16]. It is not easy to invert the Laplace transform in Eq. (28) to obtain explicitly the first-passage time distribution $P(t_f, y_0)$. However, there is an alternative way to obtain $P(t_f, y_0)$. We note that $P(t_f, y_0) = -\frac{d}{dt} Q(t, y_0)\big|_{t=t_f}$ where $Q(t, y_0)$ is the probability that the process does not cross zero up to time $t$, starting at $y_0$. An exact expression of the survival probability $Q(t, y_0)$ is known, see e.g. [21, 24];

$$Q(t, y_0) = \text{erf}\left(\frac{y_0 \sqrt{\lambda} e^{-\lambda t}}{\sqrt{1 - e^{-2\lambda t}}}\right). \tag{29}$$

Differentiating Eq. (29) with respect to $t$ and setting $t = t_f$ we get

$$P(t_f, y_0) = \frac{2 y_0 \lambda^{3/2} e^{-\lambda t_f}}{\sqrt{\pi} (1 - e^{-2\lambda t_f})^{3/2}} \exp\left\{-\frac{\lambda y_0^2 e^{-2\lambda t_f}}{(1 - e^{-2\lambda t_f})}\right\}. \tag{30}$$

It is amusing, but rather hard to see immediately, that the Laplace transform of Eq. (30) is indeed given by Eq. (28). It follows from Eq. (30) that the first-passage time distribution has an exponential tail,

$$P(t_f, y_0) \approx \frac{2}{\sqrt{\pi}} y_0 \lambda^{3/2} \exp(-\lambda t_f). \tag{31}$$

for $t_f \gg 1/\lambda$. Note that the decay coefficient $\lambda$ characterizing the exponential tail can also be computed directly from the Laplace transform in Eq. (28). Thus, since one can show that $D_{-s/\lambda}(0) = \sqrt{\pi} 2^{s/2\lambda} / \Gamma(\frac{s}{2\lambda} + \frac{1}{2})$, it follows that the Laplace transform in Eq. (28) has poles at $s = -(2m+1)\lambda$ where $m = 0, 1, 2...$. The smallest negative pole (corresponding to $m = 0$) is at $s = -\lambda$, which indicates that asymptotically for large $t_f$, the distribution $P(t_f, y_0) \sim \exp(-\lambda t_f)$.

We next compute the distribution of the area $A = \int_0^{t_f} y(t') dt'$ swept out by the process till its first-passage time by choosing $V[y] = y$ so that $T = A$. Then Eq. (32) with $V[y_0] = y_0$ can also be solved explicitly,

$$\tilde{P}(s, y_0) = e^{\lambda y_0^2/2} \frac{D_{s^2/2\lambda^3}(\sqrt{2\lambda}(y_0 + s\lambda^{-2}))}{D_{s^2/2\lambda^3}(\sqrt{2}s\lambda^{-3/2})}. \tag{32}$$

where $D_p(z)$ is the parabolic cylinder function. Once again, it is hard to invert the Laplace transform in Eq. (32) exactly. However, we anticipate that asymptotically the area distribution has an exponential tail

$$P(A, y_0) \sim \exp(-\gamma A). \tag{33}$$

The decay constant $\gamma$ must equal the smallest negative pole of the Laplace transform in Eq. (32). The poles of the Laplace transform occur at the zeros of the denominator in Eq. (32). Thus $\gamma$ is the smallest positive root of the equation

$$D_{\gamma^2/2\lambda^3}(-\sqrt{2}\gamma\lambda^{-3/2}) = 0. \tag{34}$$

Hence, the scaled variable $\phi = \gamma\lambda^{-3/2}/\sqrt{2}$ satisfies the equation $D_{\phi^2}(-2\phi) = 0$. Numerically, we find the smallest root, $\phi = 0.57279...$ which determines $\gamma$ exactly,

$$\gamma = \sqrt{2}\phi\lambda^{3/2} = 0.810048...\lambda^{3/2}. \tag{35}$$

It is difficult to determine the pre-factor in Eq. (33). However, knowledge of Eq. (35) is sufficient for the present purpose.

### B. The relationship between the discrete sequence and the continuous OU process

In this subsection, we show that as one approaches the elastic limit, $r \to 1$, the discrete sequence in Eq. (2) can be approximated by the continuous OU process in Eq. (26) with $\lambda = 1$. This is subtle, since the elastic limit itself, when $r = 1$, corresponds to Eq. (26) with $\lambda = 0$. To see what is going on, we first note from Eq. (2) that as the process evolves one can derive the general result

$$\langle u_n^2 \rangle = 1 + r^2 \langle u_{n-1}^2 \rangle = \frac{1-r^{2n}}{1-r^2} + r^{2n}u_0^2. \tag{36}$$

Therefore, for $r < 1$ and as $n \to \infty$, the velocity distribution tends to evolve towards a stationary distribution characterised by a second moment which also approaches its stationary value, $\langle u_n^2 \rangle \to 1/(1-r^2)$. Thus in the limit $\varepsilon = 1-r \to 0$, the typical value of the velocity in its stationary state becomes large, $u_n \sim 1/\sqrt{\varepsilon}$. It also follows from Eq. (36) that the number of steps $n_0$ needed for the system to reach this stationary state diverges as $\varepsilon \to 0$, $n_0 \sim 1/\varepsilon$. This suggests that we define a scaled variable, $y_n = u_n \sqrt{\varepsilon}$ which will also be of $O(1)$ for $n \geq n_0$ for the process which is constrained to be positive up to $n_0 \gg 1$. Note that $\varepsilon$ plays a similar role as $1/n_0$ in subsection II.B. For $n > n_0$, the variable $y_n$ evolves via the recursion,

$$y_n - y_{n-1} = -\varepsilon\, y_{n-1} + \sqrt{\varepsilon}\, \eta_n \tag{37}$$

starting at $y_{n_0} = u_{n_0} \sqrt{\varepsilon}$ at $n = n_0$. The rest follows as in subsection II.B. We choose $\Delta t = \varepsilon$ which is small in the elastic limit $\varepsilon \to 0$, and hence $t = (n-n_0)\Delta t$ becomes a continuous 'time' variable. Writing $y_n \equiv y(t=(n-n_0)\Delta t)$, dividing Eq. (37) by $\Delta t = \varepsilon$ and taking the limit $\varepsilon \to 0$, we find that $y(t)$ becomes an OU process as in Eq. (26) with $\lambda = 1$ and $\xi(t) = \eta_n/\sqrt{\varepsilon} = \eta_n/\sqrt{\Delta t}$. Thus, for $n > n_0$ (where $n_0 \sim 1/\varepsilon$ is large), we expect that the results for the discrete sequence can be obtained from those of the continuous process.

For example, the probability $P(n_f, u_{n_0})$ that the ball undergoes $n_f$ flights before it collapses, for $n_f > n_0$, can be read off from the corresponding continuous result in Eqs. (30) and (31) after substituting $t_f = (n_f - n_0)\Delta t = (n_f - n_0)\varepsilon$, $y_{n_0} = u_{n_0}\sqrt{\varepsilon}$ and $\lambda = 1$. For $n_f \gg n_0$, the distribution decays exponentially,

$$P(n_f, u_{n_0}) \approx \frac{2}{\sqrt{\pi}} u_{n_0} \varepsilon^{3/2} \exp(-\varepsilon n_f). \tag{38}$$

As explained earlier, in general for any $r < 1$, we expect $P(n_f) \sim \exp[-\theta_1(r) n_f]$ for large $n_f$. The result in Eq. (38) shows that in the elastic limit $r \to 1$, $\theta_1(r) \to \varepsilon = 1 - r$ independent of the noise distribution.

In a similar way, one can find the tail of the distribution $P(\tau_c)$ of the total time $\tau_c = \frac{2}{g} \sum_{n=0}^{n_f} u_n$, elapsed till the collapse, knowing the tail of the area distribution in Eq. (33). The steps are very similar to those in subsection II.B, except for the fact that one replaces $n_0$ by $1/\varepsilon$. For example, as in Eq. (20), we find that for large $\tau_c$, $\tau_c \approx S \approx \frac{2}{g} \varepsilon^{-3/2} A$ where $A$ is the area swept out by the continuous OU process till its first-passage time. Using the asymptotic distribution of $A$ in Eq. (33) and substituting $\lambda = 1$, we then find that for large $\tau_c$

$$P(\tau_c) \sim \exp\left[-\frac{1}{\sqrt{2}} \phi g \varepsilon^{3/2} \tau_c\right] \tag{39}$$

where $\phi = 0.57279...$ is smallest positive root of $D_{\phi^2}(-2\phi) = 0$. Since, for generic $r < 1$, $P(\tau_c) \sim \exp[-\theta_2(r) \tau_c]$, we find from Eq. (39) that the decay constant $\theta_2(r)$ vanishes in the elastic limit $r \to 1$ in the following fashion,

$$\theta_2(r) \approx bg(1-r)^{3/2} \tag{40}$$

where $b = \phi/\sqrt{2} = 0.405024...$ is the smallest positive root of $D_{2b^2}(-2\sqrt{2}b) = 0$, and is a universal constant independent of the noise distribution $\rho(\eta)$.

## IV. DISCUSSION

The relevance of the analysis in this paper in the context of the wider study of inelastic collapse may be summarised thus: Inelastic collapse persists in the presence of noise with features that are broadly universal when the coefficient of restitution is close to unity. In the highly complex world of real granular materials [1-3] this gives comfort in the pursuit of simple descriptions of the way such materials behave. Although the model studied is simplistic, and may be refined in many ways, particularly with regard to the precise interaction between ball and platform when both have comparable velocities, we believe the global features observed are broadly correct and will withstand more detailed scrutiny. That is not to say that a more thorough investigation would not be welcome or desirable.

We conclude by briefly considering the distribution of the *maximum* value, $M$, attained by the process given by Eq. (2) during a first passage. This is actually quite straightforward to compute and provides a different perspective on the collapse transition. We first consider the continuous process Eq. (26) with initial condition $y(t=0) = y_0$ on the interval $[0, L]$ and with absorbing barriers at $y = 0$ and $y = L$. Let $P_0(L, y_0)$ be the probability that the process is absorbed at $y = 0$. Then $P_0(L, y_0)$ is equivalent to the probability that $y(t)$ does not reach $L$ during its first passage, i.e.

$P_0(L, y_0) = P(y_m < L | y_0)$ where $y_m = \max\{y(t): 0 \leq t < t_f\}$. It follows using a well-known result that [25]

$$\left[\frac{1}{2}\frac{\partial^2}{\partial y_0^2} - \lambda y_0 \frac{\partial}{\partial y_0}\right] P_0(L, y_0) = 0 \qquad (41)$$

with boundary conditions $P_0(L, y_0 = 0) = 1$ and $P_0(L, y_0 = L) = 0$. This equation is easily solved,

$$P_0(L, y_0) = 1 - \frac{F(y_0)}{F(L)}; \qquad F(y) \equiv \int_0^y e^{\lambda x^2} dx. \qquad (42)$$

The probability density, $P(y_m, y_0)$, for the maximum value $y_m$ is given by evaluating $P(y_m, y_0) = \partial P_0(L, y_0)/\partial L \big|_{L=y_m}$. Let us now consider the discrete problem for the elastic case with $r = 1$. Setting $\lambda = 0$ one finds that $P(y_m, y_0) = y_0 / y_m^2$ for $y_m \geq y_0$. Following the procedure described in subsection II.B., the tail of the discrete distribution is then given by $P(M) \sim M^{-2}$. For the inelastic case with $r < 1$, one sets $\lambda = 1$ and follows the procedure described in subsection III.B. The tail of the discrete distribution when $\varepsilon = 1 - r$ is small is then given by $P(M) \sim \exp(-\varepsilon M^2)$. Thus the velocity of the ball during the inelastic collapse process is unlikely to attain *at any time* a value significantly greater than $\sim 1/\sqrt{\varepsilon}$.

# APPENDIX: EXPLICIT RESULTS FOR THE EXPONENTIAL DISTRIBUTION

In this Appendix we consider the discrete sequence $u_n = r u_{n-1} + \eta_n$, with an exponential noise distribution, $\rho(\eta) = \frac{1}{2}\exp(-|\eta|)$. The aim is to present an explicit calculation of the decay constant $\theta_1(r)$, characterizing the exponential tail of the distribution of the number of flights till the collapse, $P(n_f) \sim \exp[-\theta_1(r) n_f]$ for large $n_f$. The analysis complements and adds to that provided in [18]. To proceed, let us define $P_n(u|u_0)$ to be the probability that the velocity after the $n$-th collision is $u$, given that the velocities after all the preceding $(n-1)$ collisions had been positive and that the initial velocity is $u_0$. Clearly, $P_n(u|u_0)$ satisfies the following recursion,

$$P_n(u|u_0) = \frac{1}{2}\int_0^\infty P_{n-1}(u'|u_0) e^{-|u-ru'|} du' \tag{A1}$$

starting from the initial condition, $P_0(u|u_0) = \delta(u - u_0)$. Knowing $P_n(u|u_0)$ one can then calculate $Q(n, u_0)$, the probability that there is no collapse till the $n$-th step, by integrating over all possible velocities at the $n$-th step, $Q(n, u_0) = \int_0^\infty P_n(u|u_0) du$. The distribution of the number of flights before the collapse is then simply, $P(n_f, u_0) = Q(n_f - 1, u_0) - Q(n_f, u_0)$. Thus, to derive the tail behaviour of $P(n_f, u_0)$ for large $n_f$, we just need to know the solution of Eq. (A1) for large $n$. One expects that asymptotically for large $n$, $P_n(u|u_0) \to \rho^n f(u)$ [24], where the information about $u_0$ is contained in the proportionality constant. Substituting this asymptotic form into Eq. (A1), one obtains an integral-eigenvalue equation for $f(u)$,

$$\rho f(u) = \frac{1}{2}\int_0^\infty f(u') e^{-|u-ru'|} du' \tag{A2}$$

where the eigenvalue $\rho(r)$ depends continuously on $r$.

This integral equation can be transformed, on differentiating twice with respect to $u$, into a differential equation

$$\rho\left(\frac{d^2}{du^2} - 1\right) f(u) = -\frac{1}{r} f\left(\frac{u}{r}\right) \theta(u). \tag{A3}$$

For $u < 0$, the right hand side vanishes and one easily gets the solution

$$f(u) = A e^u \tag{A4}$$

where $A$ is an arbitrary constant and we have used the physical boundary condition, $f(u \to -\infty) = 0$. For $u > 0$, the right hand side of Eq. (A3) represents a nonlocal term and the solution is nontrivial. The form of Eq. (A3), however, suggests one seeks a solution of the form $f(u) = \sum_{m=0}^\infty c_m e^{-u/r^m}$ for $u > 0$. This ansatz indeed satisfies Eq. (A3) provided the coefficient $c_m$ satisfies the recursion, $c_m = c_{m-1}/[\rho r(1 - r^{-2m})]$.

Iterating this recursion, we can express every $c_m$ with $m > 0$ in terms of a single constant $c_0$. The solution for $u > 0$ is then given by

$$f(u) = c_0 \left[ e^{-u} + \sum_{m=1}^{\infty} \left[ (-\rho r)^{-m} \prod_{k=1}^{m} \frac{r^{2k}}{(1-r^{2k})} \right] e^{-u/r^m} \right]. \tag{A5}$$

We then need to match the solutions for $u < 0$ in Eq. (A4) and $u > 0$ in Eq. (A5) at $u = 0$, i.e., the function $f(u)$ and its first derivative $df/du$ must be continuous at $u = 0$. The first condition determines $c_0$ in terms of $A$. The second condition determines $\rho(r)$ which is given by the largest positive root of the equation

$$2 + \sum_{m=1}^{\infty} (-\rho r)^{-m} (1 + r^m) \prod_{k=1}^{m} \frac{r^{2k}}{(1-r^{2k})} = 0. \tag{A6}$$

As pointed out in [18], expressions of this type may be written in an alternative form using identities known from the theory of $q$-series. Thus, one can rewrite Eq. (A6) as

$$\prod_{n=0}^{\infty} (1 - \rho^{-1} r^{2n+1}) + \prod_{n=0}^{\infty} (1 - \rho^{-1} r^{2n}) = 0. \tag{A7}$$

Using the $q$-product notation, $(t,q)_{\infty} \equiv \prod_{n=0}^{\infty} (1 - tq^n)$, this can in turn be written succinctly as

$$(\rho^{-1} r, r^2)_{\infty} + (\rho^{-1}, r^2)_{\infty} = 0. \tag{A8}$$

Clearly, when $r = 0$ the solution of Eq. (A7) is $\rho = 1/2$, which is expected since the velocities are uncorrelated and the probability of $u_n$ not changing sign is just $(1/2)^n$. More generally, it is evident by inspection of Eq. (A7) that $r \leq \rho \leq 1$, so it follows that $\rho(r=1) = 1$. For any arbitrary value $0 < r < 1$, one can easily determine $\rho(r)$ numerically. Thus, for example, one finds $\rho(0.2) = 0.588203...$, $\rho(0.4) = 0.670041...$, $\rho(0.5) = 0.712667...$, $\rho(0.6) = 0.757826...$, $\rho(0.8) = 0.860729...$ etc.

Since $P_n(u|u_0) \sim [\rho(r)]^n f(u)$ for large $n$, we get $Q(n,u_0) \sim [\rho(r)]^n$ as $n \to \infty$. This implies that the distribution of the number of flights $P(n_f, u_0) \sim \exp[-\theta_1(r) n_f]$ for large $n_f$ where

$$\theta_1(r) = -\ln[\rho(r)] \tag{A9}$$

with $\rho(r)$ obtainable from Eq. (A7) as indicated. Thus, for the special noise distribution $\rho(\eta) = \frac{1}{2}\exp(-|\eta|)$, it is possible to obtain $\theta_1(r)$ to arbitrary accuracy for all $0 \leq r \leq 1$. Since $\rho(r) \to 1$ as $r \to 1$ it follows that $\theta_1(r)$ vanishes when $r = 1$. To examine how, let us consider the function $h(t,q) \equiv (t, q^2)_\infty / (tq, q^2)_\infty$. It is easy to see by cancelling factors that the function $h(t,q)$ satisfies the functional equation $h(t,q)h(tq,q) = 1-t$. By taking logarithms and expanding $\ln h(t,q)$ as a power series in $t$ one finds a formal solution [18] from which one may show as $q \to 1^-$ that,

$$h(t,q) \sim \frac{1-t}{\sqrt{1-tq}} \left( \frac{\ln t}{2\ln q} + \frac{1}{2} \right)^{1/2} \frac{\Gamma(\frac{\ln t}{2\ln q} + \frac{1}{2})}{\Gamma(\frac{\ln t}{2\ln q} + 1)}. \tag{A10}$$

We now note that one can write Eq. (A8) as $h(\rho^{-1}, r) = -1$. By setting $t = \rho^{-1}$ and $q = r$ in Eq. (A10) and setting the right hand side of Eq. (A10) equal to $-1$, it follows that $\rho(r) \approx r + \sqrt{2/\pi}(1-r)^{3/2} + \ldots$ as $r \to 1$. By invoking Eq. (A9) it then follows in turn that $\theta_1(r) \approx 1 - r - \sqrt{2/\pi}(1-r)^{3/2} + \ldots$ as $r \to 1$. This supports the more general claim in the main text that $\theta_1(r) \approx 1 - r$ as $r \to 1$ irrespective of the noise distribution.